\documentclass[pra,aps,superscriptaddress,twocolumn,floatfix]{revtex4}
\usepackage{graphicx,float,amsmath,amssymb,mathrsfs}
\usepackage[pdftex,colorlinks]{hyperref}	

\DeclareMathOperator\erfc{erfc}

\newcommand{\avg}[1]{\left\langle#1\right\rangle}
\newcommand{\be}{\begin{equation}}
\newcommand{\ee}{\end{equation}}
\newcommand{\cs}[1]{(#1)} 

\begin{document}

\title{Full distribution of the superfluid fraction and extreme value statistics in a one dimensional disordered Bose gas}

\author{M. Albert}
\affiliation{Universit\'e C\^ote d'Azur, CNRS, Institut de Physique de Nice, France}
\author{C.~A.  M\"uller} 
\affiliation{Universit\'e C\^ote d'Azur, CNRS, Institut de Physique de Nice, France}
\affiliation{German Academy of Metrology (DAM), Bavarian State Office of Weights and Measures (LMG), D-83435 Bad Reichenhall, Germany}


\begin{abstract}
The full statistical distribution of the superfluid fraction characterizing one-dimensional Bose gases in random potentials is discussed. 
Rare configurations with extreme fluctuations of the disorder potential can fragment the condensate and reduce the superfluid fraction to zero. 
The resulting bimodal probability distribution for the superfluid fraction is calculated numerically in the quasi-1D mean-field regime of ultracold atoms in laser speckle potentials. Using extreme-value statistics, an analytical scaling of the zero-superfluid probability as function of disorder strength, disorder correlation length and system size is presented. 
It is argued that similar results can be expected for point-like impurities, and that these findings are in reach for present-day experiments.   
\end{abstract}


\maketitle

\section{Introduction}

The fate of a superfluid in the presence of disorder is a famous problem in condensed-matter physics that was brought into focus by the seminal works of Giamarchi and Schultz \cite {Giamarchi1987} and Fisher \textit{et al} \cite{Fisher1989} three decades ago. While in the clean situation an interacting fluid of bosons at zero temperature is a perfect superfluid  \cite{PenroseOnsager1956}, breaking translation invariance reduces the superfluid fraction and eventually leads to an insulating state, the Bose glass phase \cite{Fisher1989}. 
These concepts, originally pioneered with superfluid $^4$He in porous media \cite{Reppy1992}, were revived with dilute ultra cold gases in optical disorder \cite{LSP2010,Shapiro2012}. 
In the regime of weakly interacting quantum fluids, the Bogoliubov approach improved considerably our understanding of the superfluid fraction \cite{Huang1992,Giorgini1994,Lopatin2002,Astrk2002,Kobayashi2002,Paul2007,Pilati2009,Pilati2010,Gaul2011}, 
and its link with the condensate fraction \cite{Mueller2012,Mueller2015} in disorder, as well as provided various techniques to draw the phase diagram of such systems \cite{Lugan2007,Falco2009,Fontanesi2009,Fontanesi2010,Fontanesi2011}. 

When dealing with disordered systems, a vital question is whether its physical properties can be fully characterized in terms of ensemble-averaged quantities or not. 
Most of the time this is the case and one can assume, for instance, that the average superfluid fraction is a good indicator of quantum transport properties of the bulk system. 
In such a situation, typical and useful observables are Gaussian distributed with decreasing relative fluctuations as the system size increases and thus become self-averaging in the thermodynamic limit. 
However, this need not always be the case, and the full probability distribution may be required to understand the physics of the strongly disordered regime. 
One example is the superfluid-insulator transition of disorder bosons in one dimension where the superfluid fraction is governed by weak links in the picture of Josephson-junction arrays  \cite{Altman2004,Altman2008,Altman2010,Vosk2012}. 
In that case the bulk physics is no longer controlled by elementary statistical properties of the disorder such as the lowest moments, but rather by its extreme value statistics. 
Another interesting example is the critical velocity of one dimensional Bose-Einstein condensates where the breakdown of superfluidity is also driven by the extreme-value statistics of the random environment \cite{Albert2008,Hulet09,Albert2010}.

\begin{figure}
  \includegraphics[width=0.9\linewidth]{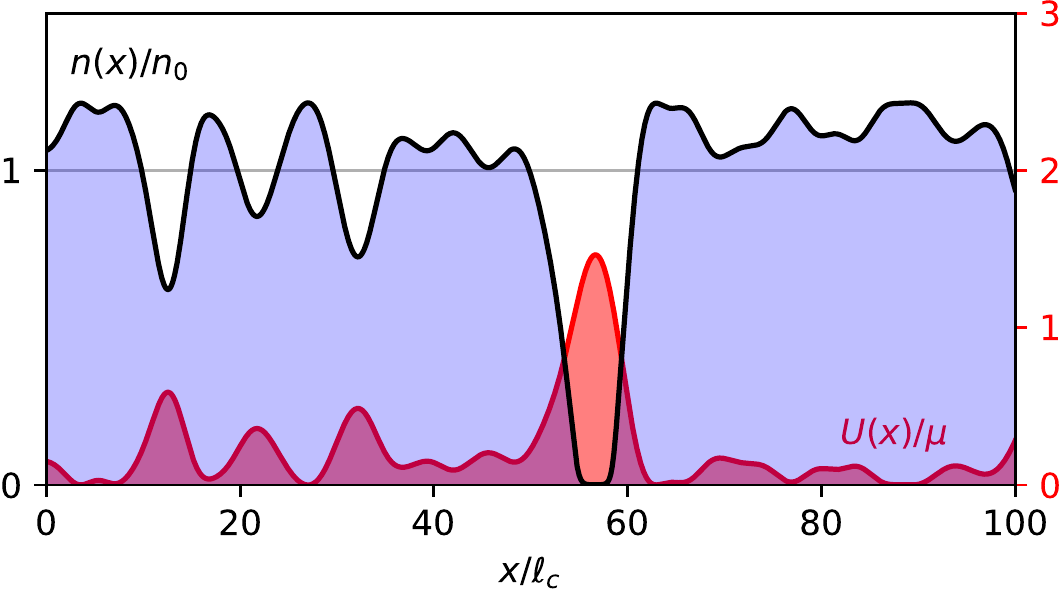}
  \caption{(Color online) Repulsive laser speckle potential $U(x)$ with average strength $\avg{U}=0.2\mu$ (bottom, red) and resulting condensate density  $n(x)$ relative to the mean density $n_0$ (top, black). 
The potential correlation length $\ell_c = 10\xi$ exceeds the condensate healing length $\xi$ such that the density is strongly suppressed by high potential peaks. 
In this particular configuration, an extreme potential fluctuation fragments the condensate, resulting in a vanishing superfluid fraction.  
  \label{fig_setup}}
\end{figure}

In this article, we discuss the statistical distribution of the superfluid fraction characterizing one-dimensional (1D) Bose-Einstein condensates (BEC) at zero temperature in a conservative disorder potential. 
For small enough and bounded disorder, the superfluid fraction becomes Gaussian distributed and self-averaging in the thermodynamic limit. 
However, unbounded disorder almost surely fragments the condensate and thus destroys superfluidity when the system size is large enough; Fig.~\ref{fig_setup} illustrates this behavior.    
Consequently, for experimentally realistic, intermediate system sizes, the full distribution of the superfluid fraction takes a bimodal form (a feature found also in the Josephson-junction model by real-space RG and quantum Monte Carlo calculations \cite{Pielawa2013}): 
A rather broad peak next to unit superfluid fraction describes standard superfluid configurations, while a rather sharp peak at zero superfluid fraction describes fragmented systems. 
In this case, obviously, the superfluid fraction is no longer well characterized by its lowest moments, mean and standard deviation, alone. 
Rather, the probability to find a fragmented instead of a superfluid configuration has to be evaluated using extreme-value statistics. 

The paper is organized as follows. 
In section \ref{sec:Model} we specify the model and describe our main qualitative observation, the rise of a bimodal superfluid distribution. 
Section \ref{sec:ncf} presents a quantitative result, namely a scaling of the normal-fraction probability with  system size, which becomes exact in the Thomas-Fermi limit.
Analogous results are expected for point-like disorder created by isolated impurities, as explained in section \ref{sec:Delta}. 
In section \ref{sec:Conclusion} we conclude and discuss an experimental strategy to observe the physics discussed along the paper. 

Additionally, appendix \ref{fs.sec} discusses a few properties of the 1D superfluid fraction, while  
appendix \ref{PerturbativeVariance.sec} contains numerical as well as perturbative analytical results for the mean superfluid fraction and its variance in spatially correlated potentials that we deem useful in the weak-disorder regime.

\section{The superfluid fraction and its probability distribution}
\label{sec:Model}

\begin{figure}
  \includegraphics[width=0.9\linewidth]{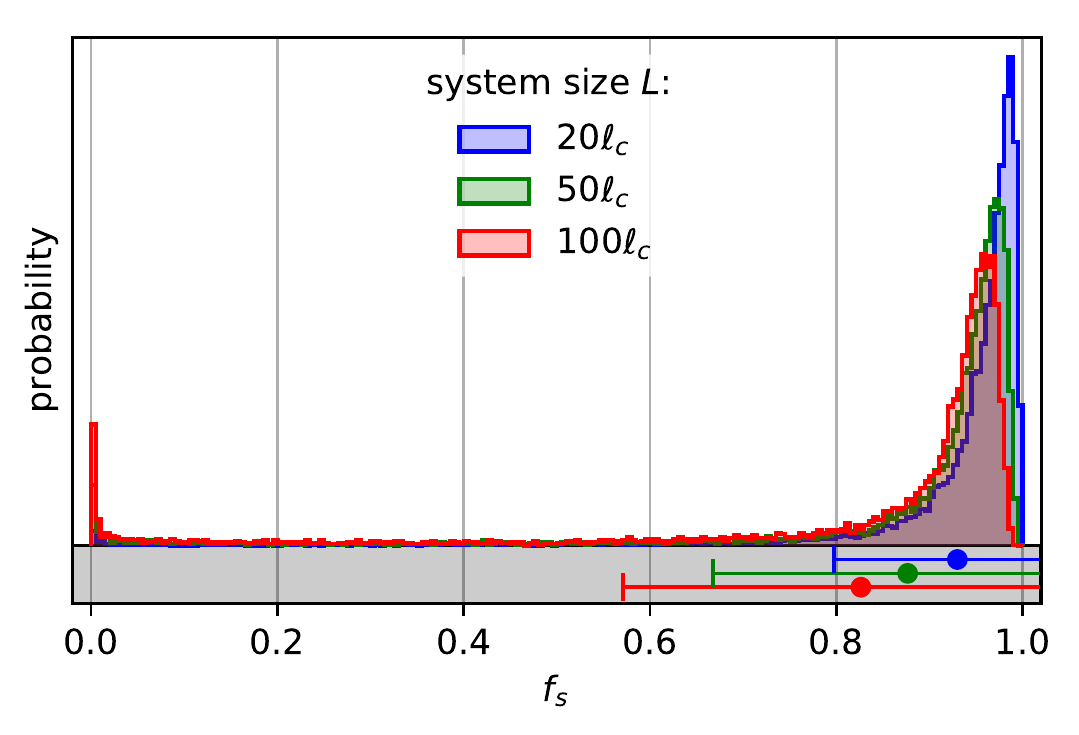}
  \caption{(Color online) 
Probability distribution of the superfluid fraction $f_s$, eq.~\eqref{fs}, for different system sizes $L$ in laser speckle disorder of strength $U_0=0.2\mu$ and correlation length $\ell_c=10\xi$ ($\mu$ and $\xi=\hbar/\sqrt{m\mu}$ are the chemical potential and healing length, respectively). 
With growing system size, the full distribution is no longer well represented by average and standard deviation, shown in the lower part of the figure. Instead, a conspicuous peak rises at $f_s=0$, signalling the occurence of extreme-value events as exemplified in Fig.~\ref{fig_setup}.       
  \label{fig_histograms}}
\end{figure}


We consider a one-dimensional Bose-Einstein condensate at rest and close to zero temperature, i.e., at temperatures low enough that thermal excitations play a negligible role. 
Certainly, at low density, quantum fluctuations destroy the phase coherence and long range order that characterize interacting Bose-Einstein condensates according to the Penrose-Onsager criterion  \cite{PenroseOnsager1956}. 
In the opposite limit of large density, transverse excitations are populated and a quasi-one dimensional description fails. 
But there is a wide range of parameters where quasi-1D mean field theory is accurate \cite{Leboeuf2001,Menotti2002,Bouchoule2011}. 
In this setting, the ground state BEC wave function $\psi(x)$ solves the Gross-Pitaevskii equation \cite{Pitaevskii2016}
\begin{equation}\label{GPeq}
\mu \psi(x)=\left[-\frac{\hbar^2}{2m}\frac{d^2}{dx^2}+U(x)+g|\psi(x)|^2\right]\psi(x). 
\end{equation}
Here, $\mu$ is the chemical potential, canonically conjugated to the number of atoms $N$ that we take to be fixed inside the system of total length $L$. $\psi$ determines the BEC density $n(x)=|\psi(x)|^2$,  $U(x)$ is a static external potential, and $g$ the contact interaction strength between atoms. 
Without an external potential, the density $n_0=N/L$ is uniform, and the chemical potential $\mu_0=gn_0$.


This work studies the impact of disorder, i.e., the effect of a random potential $U(x)$ on superfluidity; 
$\avg{\cdots}$ denotes the ensemble average over disorder configurations.
Particularly relevant for the present work, both experimentally and conceptually, 
are continuous laser speckle potentials \cite{Shapiro2012,Kuhn2007,Houches2009}. 
We focus on repulsive potentials generated by laser light that is blue-detuned from an atomic optical resonance. 
Its local potential values $U(x)$ have a one-point exponential distribution 
$p(U)= \exp[-U/U_0]\Theta(U)/U_0$ that is unbounded from above. In contrast, 
the potential values are never negative, thence the Heaviside distribution $\Theta(\cdot)$.   
At fixed number of atoms, the one-point average $\avg{U(x)} = U_0$ 
is absorbed by the chemical potential such that the relevant random process $U(x)-U_0 \mapsto U(x)$ 
has zero mean $\avg{U(x)}=0$, and the lowest possible potential value is shifted to $-U_0$. 

Laser speckle potentials, by construction from the underlying light field, are also spatially correlated. 
The spatial covariance can be written $\avg{U(x)U(x+y)}=U_0^2 C(y/\ell_c)$, 
with a correlation function $C(z)$ decaying from $C(0)=1$ to 0, and $\ell_c$ the correlation length. 
In the following, we use for definiteness a Gaussian correlation $C(z)=\exp(-z^2/2)$. 
We have checked that our results remain valid for other models of  disorder as discussed below. 
In particular, our conclusions do not depend on the precise shape of the correlation function $C(z)$ nor the on-site distribution $p(U)$, as long as the latter is unbounded, allowing for arbitrarily large, if rare, fluctuations.

The object of this work is the superfluid fraction $f_s$, the fraction of atoms supporting frictionless flow.  
Its complement, $f_n=1-f_s$, is the normal fraction that flows dissipatively; it is zero in a homogeneous BEC at zero temperature. 
Both finite temperature (by virtue of quasiparticle creation) and spatial inhomogeneity (by breaking translation invariance) create a normal component and reduce the superfluid fraction. 
As functional of the density $n(x)$ inside a quasi-1D tube of length $L$, the superfluid fraction reads \cite{Fontanesi2011,Vosk2012,Koenenberg2015} (see also App.~\ref{fs.sec})
\begin{equation}\label{fs}
f_s=\left[\frac{n_0}{L}\int_0^L \frac{dx}{n(x)}\right]^{-1}\,.
\end{equation}
The maximum value $f_s = 1$ is reached for a uniform condensate $n(x) = n_0$, and the minimum value $f_s = 0$ occurs if the density vanishes at some point, i.e., if the condensate is fragmented. Figure \ref{fig_setup} shows such a situation, resulting from the numerical solution of (\ref{GPeq}) in a particular case with a large potential peak. 

\begin{figure*}
  \includegraphics[width=0.925\linewidth]{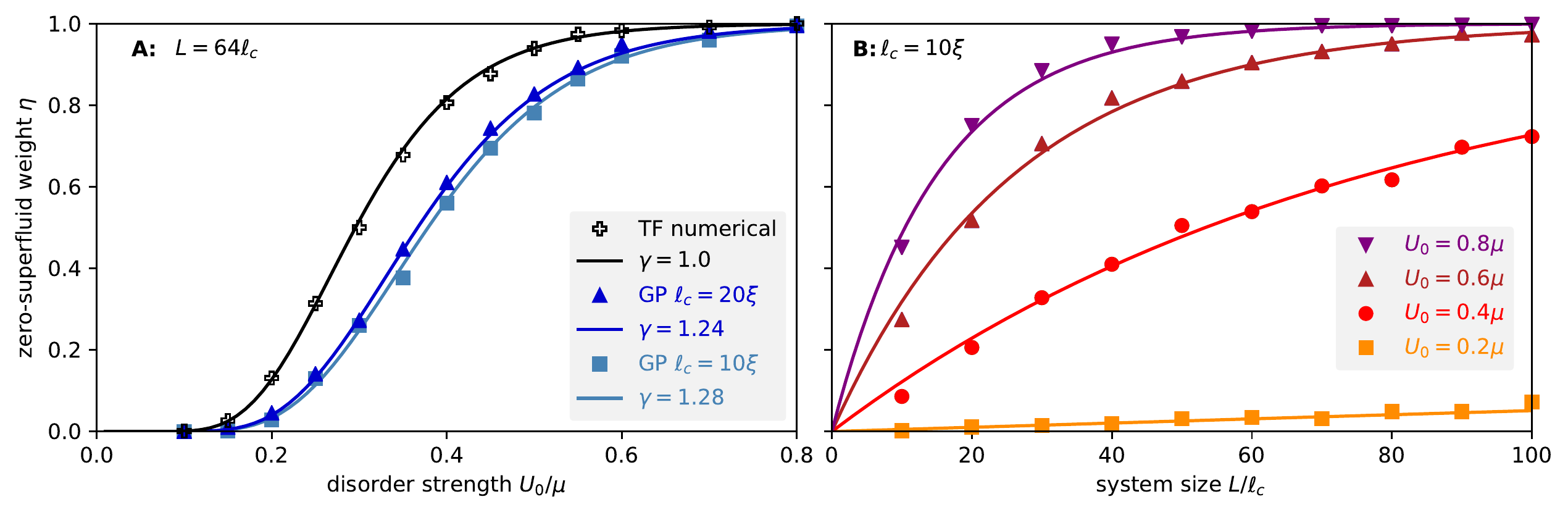}
  \caption{(Color online) Fraction $\eta$ of zero-superfluid configurations. 
 Panel \textbf{A}: $\eta$ as function of disorder strength $U_0/\mu$ for different reduced correlation lengths $\ell_c/\xi$ in systems of length $L=64\ell_c$. Open symbols are numerical results using the TF density \eqref{nTF} in \eqref{fs}, in  excellent agreement with the predicted scaling (\ref{eq_eta}) using $B=0.86L/\ell_c$ effectively iid random variables. Quantitative agreement with the GP results is reached when one accounts for smoothing of the density near its zeros by using $\mu_c= \gamma \mu$ in (\ref{eq_eta}) with $\gamma$ as noted in the legend. 
Panel \textbf{B}: $\eta$ as a function of the system length $L$ in units of the disorder correlation length $\ell_c=10\xi$ for different disorder strengths $U_0/\mu$. 
Filled symbols result from a numerical solution of the GP eq.~\eqref{GPeq} for $10^3$ disorder configurations. Full lines are the analytical prediction (\ref{eq_eta}) with $\alpha=0.86$ and $\gamma=1.28$. }  
  \label{fig_eta}
\end{figure*}

More quantitatively, Fig.~\ref{fig_histograms} shows numerically generated probability distributions for the superfluid fraction \eqref{fs} for rather weak speckle disorder of strength $U_0=0.2\mu$ and correlation length $\ell_c=10\xi$ ($\mu=\hbar^2/m\xi^2$ and $\xi$ are the bulk condensate chemical potential and healing length, respectively), for different system sizes $L$. 
Obviously, the larger the system, the higher the probability of finding a fragmented condensate, and so a conspicuous peak rises at zero superfluid fraction. 
As the probability distribution becomes bimodal, it is no longer well characterized by its mean and standard variation, displayed in the panel below the histograms. 
Instead, the probability to find fragmented condensates with zero superfluid fraction has to be evaluated using extreme-value statistics.

\section{Scaling of the zero-superfluid probability}
\label{sec:ncf}

The full probability distributions in Fig.~\ref{fig_histograms} contain a zero-superfluid peak, symbolically $p_0(f_s) = \eta \delta(f_s)$, with a weight $\eta$ that can only depend on 3 dimensionless parameters, namely disorder strength $U_0/\mu$, disorder correlation length $\ell_c/\xi$ and system size $L/\xi$.
A rather transparent functional dependence on disorder strength and system size is found in the so-called Thomas-Fermi (TF) limit $\ell_c/\xi\to \infty$ where the BEC density mirrors the external potential, 
\begin{equation}
\label{nTF} 
  n_\text{TF}(x)=\frac{\mu-U(x)}{g}\,\Theta[\mu-U(x)].
\end{equation} 
with $\Theta$ the Heaviside step function. 
This (quasi-classial) density is strictly zero at all points where the external potential exceeds the chemical potential $\mu=\mu_\text{TF}$, which needs to be tuned to ensure particle-number conservation for each realization of $U(x)$. 
In the simple TF approximation, it is rather straightforward to estimate the zero-superfluid weight $\eta$.   
Indeed, one can now link its complement $1-\eta$ to the probability that the condensate is not fragmented, i.e., that the disorder potential nowhere exceeds the chemical potential:  
$1-\eta = \textrm{Prob}(\forall x\in[0,L]:U(x)\le \mu) $. 

To progress, we approximate the smoothly correlated disorder potential by a discrete set of $B = \alpha L/\ell_c$  independent random variables $U_i$ with the same distribution $p(U)$, as discussed in \cite{Albert2010} following extreme value statistics of correlated random continuous variables \cite{Pickands1969,extremevalues}.   
The coefficient $\alpha$ of order 1 depends only weakly on the disorder distribution and will be fixed later.
The probability of all $B$ variables $U_i$ being smaller than $\mu$ then is  
\begin{equation} \label{prob_U_below_mu}
1-\eta =\textrm{Prob}(\forall i=1,\dots,B : U_i\le \mu) = P(\mu)^B, 
\end{equation} 
where $P(x)=\int_{-\infty}^{x} dU p(u)$ is the cumulative one-point distribution. 
For a blue-detuned, zero-centered speckle potential with 
$p(U) = U_0^{-1} \exp[-(U+U_0)/U_0] \Theta (U+U_0)$, the expected zero-superfluid weight then amounts to
\begin{equation}   
\label{eq_eta}
  \eta=1-\left[1-\exp(-1-\mu/U_0)\right]^B\,.
\end{equation}
The numerical prefactor in the number $B=\alpha L/\ell_c$ of iid variables can be fixed by fitting this prediction to the result of a numerical calculation using the TF density \eqref{nTF} in Eq.~\eqref{fs} for various values of $U_0/\mu$  and $L/\ell_c$

Figure \ref{fig_eta} panel A shows excellent agreement between this analytical prediction and numerical TF results for $\alpha=0.86$. 
Quantitative agreement is reached with the GP results (full symbols) when taking into account the smoothing of the GP density compared to the TF approximation.  
Indeed, for finite values of $\ell_c/\xi$, the TF density is too rough an approximation to describe the fine details of the density near its zeros where quantum corrections induce finite, if small densities even the classically forbidden regions, and thus cannot be expected to give the 
superfluid fraction with quantitative precision. 
By using $\mu_c= \gamma \mu$ with $\gamma$ of order unity as a slightly larger critical value for the threshold in Eqs.~\eqref{prob_U_below_mu} and \eqref{eq_eta}, we find excellent agreement also between the GP data and the scaling \eqref{eq_eta}, as shown in both panels of Fig.~\ref{fig_eta}. 
The TF limit with $\gamma=1$ is (slowly) reached as the ratio $\ell_c/\xi$ increases. 
Independently of the numerical fit quality, the extreme-value statistics argument behind Eq.~\eqref{eq_eta} essentially captures the physics of the zero-superfluid weight. 
Also, we have checked that analogous results apply to various local distributions $p(U)$ and correlation functions, as long as $p(U)$ is unbounded and correlations decay faster than a logarithm \cite{Albert2010,Pickands1969}.

\section{Point-like impurities} 
\label{sec:Delta}

When the correlation length is reduced, away from the TF limit and toward the uncorrelated-disorder limit, the screening of disorder by interaction could be expected to minimize the extreme-value effects and lead again to a self-averaging situation. 
However, the extreme-value argument stays valid and still describes the destruction of superfluidity in the thermodynamic limit, as we show in this section in the extreme opposite case of completely uncorrelated disorder.  
We start with a model of point-like impurities:   
\begin{equation}\label{Uimp}
  U_\delta(x)=\lambda \mu \xi\sum_{i=1}^{M} \delta (x-x_i) .
\end{equation}
The parameter $\lambda$ describes each impurity's strength relative to the chemical potential. 
The positions $x_i$ are iid random variables uniformly distributed in $[0,L]$ with density $\nu=M/L$ taken to be constant in the thermodynamic limit $L,M\to \infty$.  
Such a potential is uncorrelated, with covariance 
$\avg{U_\delta(x) U_\delta(y)}-\avg{U_\delta}^2=(\lambda\mu\xi)^2 \nu \delta (x-y)$.  

Let us first calculate the disorder-induced normal fraction in the weak disorder limit, where perturbation theory  \cite{Paul2007} yields   
\begin{equation}\label{rhop}
  f_n=\frac{1}{2\xi \mu^2 L}\int_{[0,L]^2} dx dy\,U(x)U(y) \mathcal K(x-y)\,
\end{equation}
with $\mathcal K(z)=(1+2|z|/\xi)\,e^{-2|z|/\xi}$. 
For the point-like impurities \eqref{Uimp} this reduces to 
\begin{equation}
   f_n=\frac{\lambda^2\xi}{2L}\sum_{i,j=1}^M\mathcal K(x_i-x_j)\,.
\end{equation}
In the scarce-impurity limit $\xi \nu\ll 1$, the dominant part comes from the diagonal terms $i=j$ and one recovers the result $\avg{f_s}=1-\frac{1}{2}\lambda^2\xi \nu$ of Huang and Meng \cite{Huang1992} for the thermodynamic limit. 

It is instructive to look at a large, but finite system of length $L\gg\xi$. 
The expectation value of the superfluid fraction is 
$\avg{f_s}= 1- \frac{1}{2}\lambda^2\xi \nu \left[1- 2 \xi \nu + 2 \xi/L\right]$, and its variance $\Delta f_s^2=\frac{5}{8} 
(\lambda^2\xi \nu)^2  \xi/L$. 
Hence the fluctuations are predicted to decay as $1/L$ in seeming accordance with the central limit theorem, such that $f_s$ would be self averaging. 

However, this perturbative prediction neglects rather improbable, but highly relevant disorder configurations where several impurities cluster together.  
Indeed, when distributed independently, impurities can be located in close vicinity, combining their strengths and strongly depleting both the local density and superfluidity \cite{Albert2010}. 
As the system size grows, impurity clusters that are large enough to fragment the condensate become increasingly likely, thereby contributing to the zero-superfluid weight. 
Obviously such a situation is beyond perturbation theory, and more sophisticated techniques involving extreme-value statistics are required.

Qualitatively, the argument runs as follows:       
If $k$ impurities cluster within a healing length $\xi$ or less, the condensate effectively sees a single impurity of strength $k\lambda$. 
Divide then the disordered region into $B=L/\xi$ boxes. The condensate will not be fragmented 
if the maximum number of impurities inside each box, $K=\max \{k_1,k_2,...,k_B\}$, 
is smaller than a certain critical value $K_c=\gamma/\lambda$. 
Here $\gamma$ is a number of order unity that depends on the threshold below which $f_s$ is counted as zero. 
Based on the single-impurity problem, one can estimate $\gamma$ to be around four or five in order to have $f_s\le 0.05$.  
In the framework of this simple picture, that was proven to be accurate \cite{Albert2010}, the probability of finding $k$ out of $M$ impurities in any box is $\pi(k)= \binom{M}{k} p^{k} (1-p)^{M-k}$, where $p=1/B$. In the limit of a wide disordered region ($L\to\infty$), the product $pM=\xi \nu=:\zeta$ remaining constant, this binomial law can be approximated by a Poisson law: $\pi(k)\simeq e^{-\zeta}\,\zeta^{k}/k!$. 
In this limit the variables $k_i$ are uncorrelated, and (compare with Eq.~\eqref{prob_U_below_mu})     
\begin{equation} 
\textrm{Prob}(\forall i=1,\dots,B : k_i \leq K_c) = \Pi (K_c)^B
\end{equation}
where $\Pi(K)=\Gamma(K+1,\zeta)/K!$ is the cumulative Poisson distribution, with $\Gamma(x,\zeta)$ the incomplete gamma function. 
The fraction of disorder realizations where the condensate is fragmented and no longer superfluid thus is the complement
\begin{equation}\label{eq_eta2}
\eta=1-\Pi(K_c)^B.
\end{equation}
It must be kept in mind, however, that the typical value of $K$ grows very slowly with $L$ (typically logarithmically), so that strong effects of impurity clusters can only be observed in very large systems. 
For instance, in order to find $\eta=0.1$ with $\xi \nu=0.3$ and $\lambda=0.8$, Eq.~(\ref{eq_eta2}) requires a system of size $L\simeq 1.5\cdot10^5\xi$, which is out of reach for our current numerical calculations.  
Nevertheless, point-like impurities have the same qualitative effect on superfluid fraction statistics as a smooth speckle potential.

\section{Conclusion}
\label{sec:Conclusion}
In the 1d-mean field regime, we have evaluated the impact of random potentials on the full statistical distribution of the superfluid fraction $f_s$. 
As the system size $L$ grows, large fluctuations of an unbounded potential like laser speckle or clusters of impurities become more and more probable and eventually fragment the BEC. In such a situation, the full probability distribution of $f_s$ is bimodal and the mean superfluid fraction is no longer the only relevant quantity characterizing the physical properties. A peak at $ f_s=0$ develops as the hallmark of fragmentation and grows with the system size. 

This result is of course peculiar to one dimensional systems. 
While in one dimension there is no BEC in the thermodynamic limit due to quantum fluctuations \cite{Pitaevskii2016}, a proper analysis of these fluctuations \cite{Petrov2000} shows that the coherence length of the quasi condensate is generally larger than a few centimeters whereas the typical size of an atomic BEC is less than a few hundred microns and therefore phase coherence is preserved in disorder as it has been demonstrated experimentally \cite{Hulet09}. 
Superfluidity can then be destroyed before Bose-Einstein condensation. 
For a speckle potential the correlation length can easily be tuned to be of the order of one or several micrometers and the healing length around 0.2$\,\mu$m. If the cloud size is of about a few hundred microns, the situation  described in this paper is easily within reach. 

The one-dimensional Gross-Pitaevskii equation is only a relatively simple mean-field approximation of a quasi-1d Bose-gas. 
Although most of the results presented in this paper should be qualitatively correct, one may wonder about their validity in the presence of quantum fluctuations and transverse degrees of freedom.  
It is most likely that quantum fluctuation will not help to preserve superfluidity but will certainly affect quantitatively the results as they become more and more important for instance away from the weakly interacting limit. 
Moreover, the physics discussed in this work being purely one dimensional, it would be important to understand how the results are affected in the one dimensional to three dimensional cross over even in the weakly interacting limit. We leave these interesting question for further research. 

\begin{acknowledgments} We acknowledge fruitful discussions with P.E.~Larr\'e, N.~Pavloff and P.~Vignolo. This paper is dedicated to the memory of Patricio Leboeuf, whose kindness, trust and freedom offered to M.A. while being his PhD student at LPTMS will always be gratefully remembered. 
\end{acknowledgments}  

\appendix

\section{Superfluid fraction in 1D}
\label{fs.sec}

Let $\phi(x)=\sqrt{n(x)} e^{i\varphi(x)}$ be the mean-field order parameter of a 1D Bose gas, with $n(x)=|\phi(x)|^2$ the stationary condensate density, and $\varphi(x)$ a local phase. The phase gradient determines the superfluid velocity to $v(x) = (\hbar/m)\partial_x\varphi(x)$. The superfluid current density is $j(x)=n(x)v(x)$. This current density is actually independent of position $x$ because of the continuity equation (mass conservation) $\partial_t n +\partial_x j=0$, such that $v(x) = j/n(x)$ everywhere.  

Consider now a 1D section of finite length $L$; the total phase twist accumulated from left to right is 
\be\label{Delphi}
\Delta\varphi = \int_0^L dx \partial_x\varphi(x) = \frac{m}{\hbar} \int_0^L dx \frac{j}{n(x)}.
\ee
In the limit $\Delta\varphi\to0$, the proportionality factor between phase twist and current
defines the superfluid density $n_s$: 
\be\label{jnsDelphi}
j = n_s \frac{\hbar \Delta\varphi }{Lm}.  
\ee  
The two identities \eqref{Delphi} and \eqref{jnsDelphi} determine the inverse superfluid density to 
\be
n_s^{-1} = \int_0^L \frac{dx}{L} \frac{1}{n(x)}.
\ee 
At fixed total atom number $N=\int_0^L dx n(x) = Ln_0$, the inverse superfluid fraction $f_s^{-1}=n_0/n_s$ then is \be\label{fsdef}
f_s^{-1} =  \int_0^L \frac{dx}{L} \frac{n_0}{n(x)}. 
\ee 
Various derivations of this result have been published  
\cite{Fontanesi2011,Vosk2012,Koenenberg2015}, none of them quite as short or elementary, it seems.

The superfluid fraction is bounded by $f_s\leq 1$ because it is the continuum limit of the harmonic mean of random variables $y_i=n(x_i)/n_0$ at discrete points $x_i$ \cite{Fontanesi2010}. 
The bound $f_s\leq 1$ also follows from the Cauchy-Schwartz inequality $\cs{f|g} ^2 \leq \cs{f|f}\cs{g|g}$ for the $L_2([0,L])$ scalar product $\cs{f|g}=L^{-1}\int_0^L dx f(x) g(x)$ by using $f(x)=\sqrt{n_0/n(x)}$ and $g(x)=f(x)^{-1}=\sqrt{n(x)/n_0 }$ such that $\cs{f|f}=f_s^{-1}$ while $\cs{g|g} = 1 = \cs{f|g}$.

The maximum value $f_s=1$ is obtained for a uniform condensate ($n(x)=n_0$), and the minimum value $f_s=0$ occurs if the density vanishes at some point at least as $n(x) \sim |x-x_0|^p,p\geq 1$, i.e., if the condensate is fragmented.

\section{Average superfluid fraction and its variance for weak disorder}
\label{PerturbativeVariance.sec}

\begin{figure}
  \includegraphics[width=0.95\linewidth]{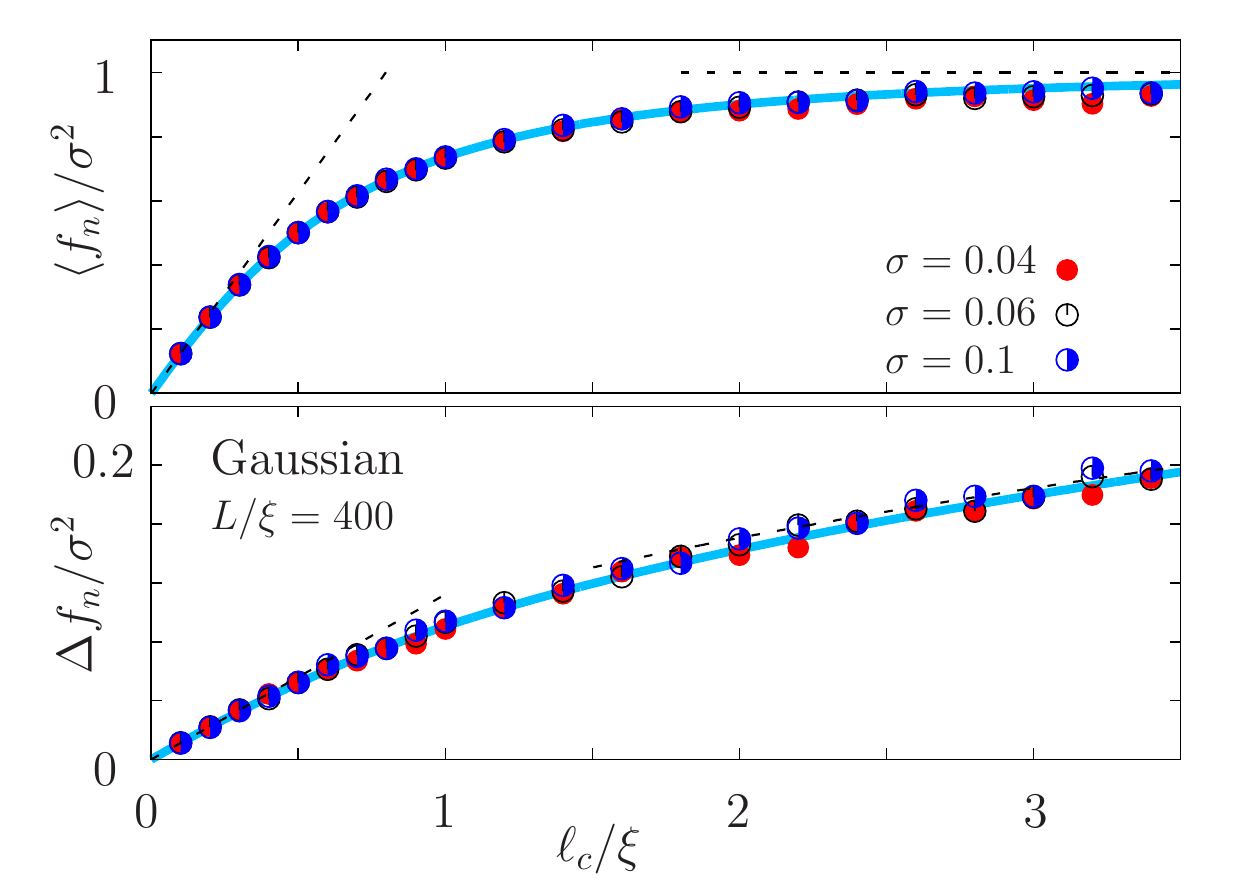}
  \caption{(Color online) Mean normal fraction $\avg{f_n}$ and its standard deviation $\Delta f_n$ of a quasi 1D condensate in a  Gaussian potential of length $L=400\xi$ ($\xi$ is the healing length),  
 plotted as function of the reduced correlation length $z_c=\ell_c/\xi$ for various disorder strengths $\sigma=U_0/\mu$. The continuous blue lines are the analytical results (\ref{rhop2}) and \eqref{drhop2}, respectively, and the dashed lines are the associated asymptotic expressions (see main text).}
  \label{fig_rhon}
\end{figure}

\begin{figure}
  \includegraphics[width=0.95\linewidth]{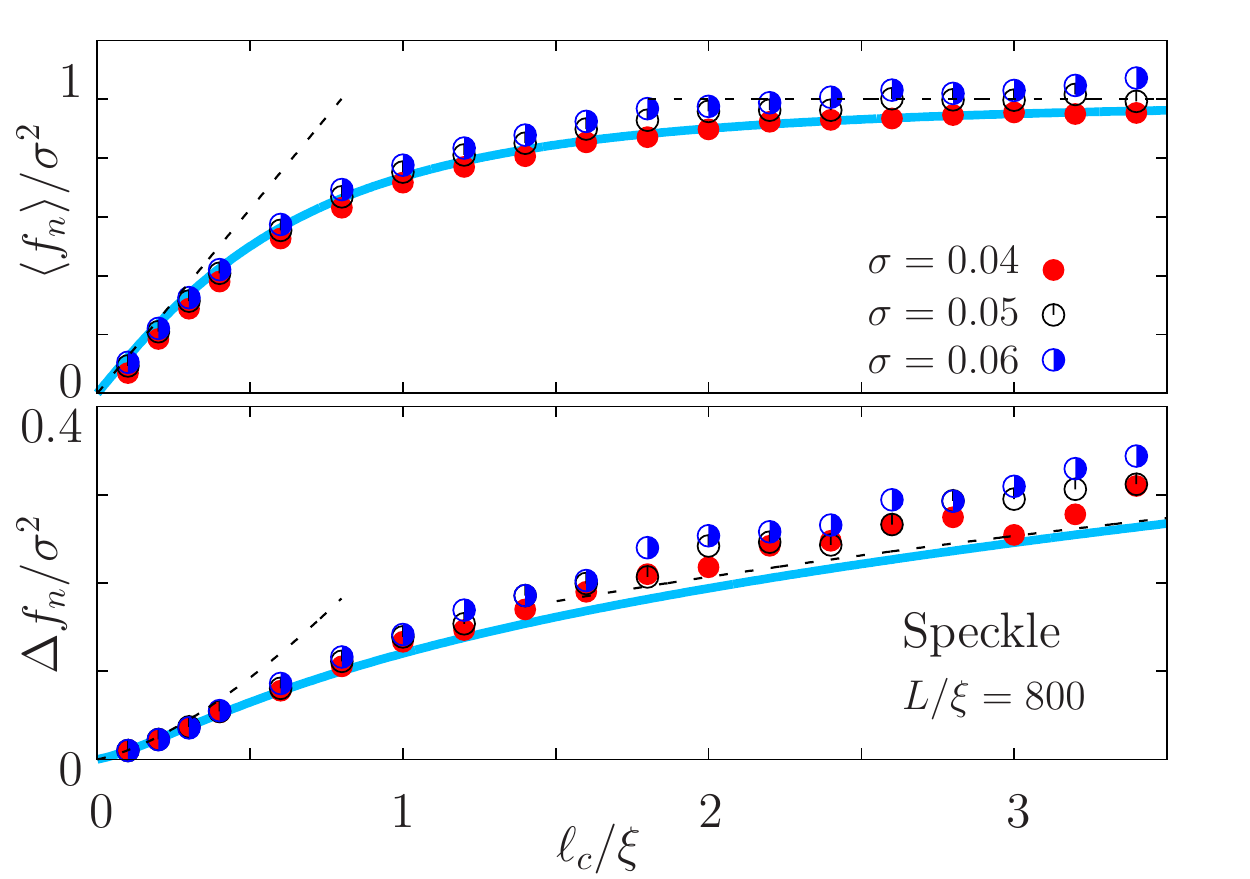}
  \caption{(Color online) Mean normal fraction $\avg{f_n}$ and standard deviation $\Delta  f_n$ as a function of the correlation length $\ell_c$ for various values of the disorder strength $\sigma$ of the speckle potential of length $L=800\xi$ ($\xi$ is the healing length). The continuous blue line in the upper panel is the analytical result (\ref{rhop2}) and the dashed lines are asymptotic expressions (see main text).}
  \label{fig_rhonS}
\end{figure}

\subsection{Gaussian random process, Gaussian correlation}

Within the perturbative regime, it is possible to obtain analytical results for Gaussian correlated, Gaussian random processes in the interesting limit of large systems $L\gg\ell_c,\xi$. 
Using $\avg{U(x)U(x+y)}=U_0^2 \exp[-y^2/(2\ell_c^2)]$ in Eq.~\eqref{rhop}, 
the mean normal fraction reads 
\begin{equation}\label{rhop2}
  \avg{f_n}=\sigma^2 z_c \left[2z_c+\sqrt{\tfrac{\pi}{2}}(1-4z_c^2) e^{2z_c^2} \erfc(\sqrt{2}z_c)\right],
\end{equation}
where $z_c=\ell_c/\xi$ and $\sigma=U_0/\mu$. 
In the TF limit $z_c\gg1$ of slowly varying potentials it reduces to $\avg{ f_n}=\sigma^2$. 
In the white noise limit $z_c\ll 1$, it is $\avg{ f_n}=\sigma^2\sqrt{\frac{\pi}{2}}z_c$. 
The upper panel of figure \ref{fig_rhon} compares the results of the numerical solution with the perturbative prediction (\ref{rhop2}) for various but small disorder amplitudes. 
The agreement is very good as expected in this regime.

Assuming a Gaussian process one can also compute the variance of $ f_n$ (which is equal to the variance of $f_s$), using Wick's theorem, 
\begin{eqnarray}\label{Wick}
\avg{U_1 U_2 U_3 U_4}_\text{Gauss} &=& \avg{U_1 U_2}\avg{U_3 U_4}+\avg{U_1 U_3}\avg{U_2 U_4} \nonumber \\ 
 && + \avg{U_1 U_4}\avg{U_2 U_3} 
\end{eqnarray} 
where $U_i=U(x_i)$. 
This yields
\begin{eqnarray} \label{drhop2}
  \Delta f_n^2 & = &\frac{\sigma^4\,\xi}{12 L}\,z_c^2\,\Big \{4\sqrt{\pi}z_c(15-32z_c^2+64z_c^4)  \\ 
  &&  
   +\pi e^{4z_c^2}\,[15-8z_c^2(9-24z_c^2+64z_c^2)]\,\textrm{erfc}(2 z_c)\Big\}\, ,\nonumber
\end{eqnarray}
 which simplifies for $l_c\ll \xi $ to $\Delta f_n^2=\frac{5\pi}{4}\frac{\ell_c^2\sigma^4}{L\xi}$ 
 and for $\ell_c\gg \xi$
to $\Delta f_n^2=\frac{2\sqrt{\pi}\sigma^4\ell_c}{L}$. 
These results are compared to numerical calculations in the lower panel of Fig. \ref{fig_rhon}, again with excellent agreement. 

Thus, at the level of perturbation theory, the standard deviation of $f_s$ scales as $1/\sqrt{L}$, which suggests that the superfluid fraction is indeed a self averaging quantity, which then should, by virtue of the central limit theorem, be Gaussian distributed in a large enough system.  
However, as pointed out in the main part of the paper, in large enough systems, extreme events fragment the condensate and thus induce a zero-superfluid weight in the full probability distribution that cannot be accounted for  by perturbation theory.

\subsection{Laser speckle potential, Gaussian correlation}

For Gaussian-correlated, zero-centered laser speckle, the mean normal fraction is also given by 
\eqref{rhop2}. For the variance, however, we expect differences because only the electric field amplitude of fully developed laser speckle is a Gaussian random process. 
The optical potential acting on the atoms is proportional to the intensity, and therefore 
corrections to Wick's theorem for higher than second-order moments have to be included \cite{Kuhn2007,Houches2009}. 
For the expectation value of a product of 4 potential values $U_i=U(x_i)=|E_i|^2$, one has
\begin{equation}
\begin{split} 
    &\avg{ U_1 U_2 U_3 U_4}_\text{Speckle} = \avg{U_1 U_2 U_3 U_4}_\text{Gauss} \\ 
    &+ 2\textrm{Re} \big\{\gamma_{12}\gamma_{23}\gamma_{34}\gamma_{41}
    + \gamma_{12}\gamma_{24}\gamma_{43}\gamma_{31}+\gamma_{13}\gamma_{32}\gamma_{24}\gamma_{41}\big\},
\end{split}
\end{equation}
where $\gamma_{ij}=\avg{E_i^*E_j}=U_0\exp[-(x_i-x_j)^2/4l_c^2]$.
Inserting this in the perturbative expression (\ref{rhop}) allows to compute the standard deviation of the superfluid fraction for such a potential. We have not found a closed-form expression for all contributions, but the agreement 
between the perturbative calculation and the numerical data displayed in Fig.~\ref{fig_rhonS} is quite satisfactory. 
In the TF limit $\ell_c \gg \xi$, one finds $\Delta f_n^2=\frac{\sigma^4}{L}\, 4\sqrt{\pi}\ell_c (1+\sqrt{2})$  and in the white-noise limit $\ell_c\ll \xi$ one finds $\Delta f_n^2=\frac{\sigma^4}{ L\xi}\,\ell_c^2\left\{\frac{5\pi}{4}+3(2\pi)^{3/2} \,\ell_c/\xi\right\}$.


\begin{thebibliography}{99}

\bibitem{Giamarchi1987} T. Giamarchi, and H.~J. Schultz, Eur. Phys. Lett. {\bf 3}, 1287 (1987).
\bibitem{Fisher1989} M.~P.~A. Fisher, P.~B. Weichman, G. Grinstein, and D.~S. Fisher, Phys. Rev. B {\bf 40}, 546 (1989).

\bibitem{PenroseOnsager1956} 
O.~Penrose and L.~Onsager, Phys. Rev. \textbf{104}, 576 (1956).
\bibitem{Reppy1992} J.~D. Reppy, J. Low Temp. {\bf 87}, 205 (1992).
\bibitem{Shapiro2012} B. Shapiro, J. Phys. A: Math. Theor. {\bf 45}, 143001 (2012).
\bibitem{LSP2010} L. Sanchez-Palencia, and M. Lewenstein, Nature Physics {\bf 6}, 87-95 (2010).

\bibitem{Huang1992} K. Huang and H.-F. Meng, Phys. Rev. Lett. \textbf{69}, 644 (1992).
\bibitem{Giorgini1994} S. Giorgini, L. Pitaevskii, and S. Stringari, Phys. Rev. B {\bf 49}, 12938 (1994).
\bibitem{Lopatin2002} A.~V. Lopatin, and V.~M. Vinokur, Phys. Rev. Lett. {\bf 88}, 235503 (2002).
\bibitem{Astrk2002} G.~E. Astrakharchik, J. Boronat, J. Casulleras, and S. Giorgini, Phys. Rev. A {\bf 66}, 023603 (2002).
\bibitem{Kobayashi2002} M. Kobayashi, and M. Tsubota, Phys. Rev. B {\bf 66}, 174516 (2002).
\bibitem{Paul2007} T. Paul, P. Schlagheck, P. Leboeuf, and N. Pavloff, Phys. Rev. Lett. \textbf{98}, 210602 (2007).
\bibitem{Pilati2009} S. Pilati, S. Giorgini, and N. Prokof\'{ }ev, Phys. Rev. Lett. {\bf 102}, 150402 (2009).
\bibitem{Pilati2010} S. Pilati, S. Giorgini, M. Modugno, and N. Prokof\'{ }ev, New J. Phys. {\bf 12}, 073003 (2010).
\bibitem{Gaul2011} C. Gaul, and C.~A. M\"uller, Phys. Rev. A {\bf 83}, 063629 (2011).

\bibitem{Mueller2012} C.~A.~M\"uller and C.~Gaul, New J.~Phys. \textbf{14}, 075025 (2012). 
\bibitem{Mueller2015} C~A.~M\"ulller, Phys. Rev. A \textbf{91}, 023602 (2015). 

\bibitem{Lugan2007} P. Lugan, D. Cl\'ement, P. Bouyer, A. Aspect, M. Lewenstein, and L. Sanchez-Palencia, Phys. Rev. Lett. {\bf 98}, 170403 (2007).
\bibitem{Falco2009} G.~M. Falco, T. Nattermann, and V.~L. Pokrovsky, Phys. Rev. B {\bf 80}, 104515 (2009).
\bibitem{Fontanesi2009}	L. Fontanesi, M. Wouters, and V. Savona, Phys. Rev. Lett. \textbf{103}, 030403 (2009).
\bibitem{Fontanesi2010} L. Fontanesi, M. Wouters, and V. Savona, Phys. Rev. A \textbf{81}, 053603 (2010).
\bibitem{Fontanesi2011} L. Fontanesi, M. Wouters, and V. Savona, Phys. Rev. A \textbf{83}, 033626 (2011).


\bibitem{Altman2004} E. Altman, Y. Kafri, A. Polkovnikov, and G. Refael, Phys. Rev. Lett. {\bf 93}, 150402 (2004).
\bibitem{Altman2008} E. Altman, Y. Kafri, A. Polkovnikov, and G. Refael, Phys. Rev. Lett. {\bf 100}, 170402 (2008).
\bibitem{Altman2010} E. Altman, Y. Kafri, A. Polkovnikov, and G. Refael, Phys. Rev B {\bf 81}, 174528 (2010).
\bibitem{Vosk2012} R. Vosk, and E. Altman, Phys. Rev. B {\bf 85}, 024531 (2012).


\bibitem{Albert2008} M. Albert, T. Paul, N. Pavloff, and P. Leboeuf, Phys. Rev. Lett \textbf{100}, 250405 (2008).
\bibitem{Hulet09} D. Dries, S. E. Pollack, J. M. Hitchcock, and R. G. Hulet, Phys. Rev. A {\bf 82}, 033603 (2010).
\bibitem{Albert2010} M. Albert, T. Paul, N. Pavloff, and P. Leboeuf, Phys. Rev. A \textbf{82}, 011602(R) (2010). 
\bibitem{Pielawa2013} S.~Pielawa and E.~Altman, Phys. Rev. B \textbf{88}, 224201 (2013). 

\bibitem{Leboeuf2001} P. Leboeuf and N. Pavloff, Phys. Rev. A {\bf 64}, 033602 (2001).
\bibitem{Menotti2002} C. Menotti and S. Stringari, Phys. Rev. A 66, 043610 (2002).
\bibitem{Bouchoule2011} I. Bouchoule, N. J. Van Druten, and C. I. Westbrook, in Atom Chips, edited by J. Reichel and V. Vuleti\'c (Wiley Online Library, 2011).
\bibitem{Pitaevskii2016} L. Pitaevskii and S. Stringari, \textit{Bose-Einstein Condensation and Superfluidity} (Oxford University Press, Oxford, UK, 2016).

\bibitem{Kuhn2007} R.\,C.~Kuhn, O.~Sigwarth, C.~Miniatura, D.~Delande, C.\,A.~M\"uller, 
New J. Phys. \textbf{9}, 161 (2007).
\bibitem{Houches2009} C. A. M\"uller and D. Delande, \textit{Lecture notes Les Houches School of Physics on "Ultracold gases and quantum information" 2009 in Singapore}, edited by C. Miniatura \textit{et al} (Oxford University Press, 2011).
\bibitem{Koenenberg2015}
M. K\"onenberg, T. Moser, R. Seiringer, J. Yngvason, New J. Phys. \textbf{17}, 013022 (2015). 

\bibitem{Pickands1969} J. Pickands, Trans. Amer. Math. Soc. \textbf{145}, 75 (1969).
\bibitem{extremevalues} E. J. Gumbel, {\it Statistics of Extremes} 
(Dover Publications, New York, 2004).

\bibitem{Petrov2000} D.~S. Petrov, G.~V. Shlyapnikov, and J.~T.~M. Walraven, Phys. Rev. Lett. {\bf 85}, 3745 (2000).


\end{thebibliography}
\end{document}